\documentclass[twocolumn,prl,superscriptaddress]{revtex4}
\usepackage{amsfonts}
\usepackage{amssymb}
\usepackage{amsmath}
\usepackage{epsfig}
\usepackage{color}
\usepackage{graphics, graphicx}
\usepackage{bbold}
\usepackage{psfrag}
\usepackage{mathcomp}
\usepackage{subfigure}
\usepackage{verbatim}
\usepackage[colorlinks,citecolor=blue]{hyperref}

\makeatletter

\newcommand{\Rmnum}[1]{\expandafter\@slowromancap\romannumeral #1@}
\makeatother

\begin{document}

\title{Digital Simulation of Topological Matter on Programmable Quantum Processors}

\author{Feng Mei}
\email{meifeng@sxu.edu.cn}
\affiliation{State Key Laboratory of Quantum Optics and Quantum Optics Devices, Institute
of Laser Spectroscopy, Shanxi University, Taiyuan, Shanxi 030006, China}
\affiliation{Collaborative Innovation Center of Extreme Optics, Shanxi
University, Taiyuan, Shanxi 030006, China}

\author{Qihao Guo}
\affiliation{School of Science, Xian Jiaotong University,
Xian 710049, Shaanxi, China}

\author{Ya-Fei Yu}
\affiliation{Guangdong Provincial Key Laboratory of Quantum Engineering and Quantum Materials, School of Information and Optoelectronic Science and Engineering, South China Normal University, Guangzhou 510006, China}

\author{Liantuan Xiao}
\affiliation{State Key Laboratory of Quantum Optics and Quantum Optics Devices, Institute
of Laser Spectroscopy, Shanxi University, Taiyuan, Shanxi 030006, China}
\affiliation{Collaborative Innovation Center of Extreme Optics, Shanxi
University, Taiyuan, Shanxi 030006, China}

\author{Shi-Liang Zhu}
\email{slzhu@nju.edu.cn}
\affiliation{National Laboratory of Solid State Microstructures, School of Physics,
Nanjing University, Nanjing 210093, China}
\affiliation{Guangdong Provincial Key Laboratory of Quantum Engineering and Quantum
Materials, GPETR Center for Quantum Precision Measurement, Frontier Research Institute for Physics and SPTE, South China Normal University, Guangzhou 510006, China}

\author{Suotang Jia}
\affiliation{State Key Laboratory of Quantum Optics and Quantum Optics Devices, Institute
of Laser Spectroscopy, Shanxi University, Taiyuan, Shanxi 030006, China}
\affiliation{Collaborative Innovation Center of Extreme Optics, Shanxi
University, Taiyuan, Shanxi 030006, China}

\date{\today}

\begin{abstract}
Simulating the topological phases of matter in synthetic quantum simulators is a topic of considerable interest. Given the universality of digital quantum simulators, the prospect of digitally simulating exotic topological phases is greatly enhanced. However, it is still an open question how to realize digital quantum simulation of topological phases of matter. Here, using common single- and two-qubit elementary quantum gates, we propose and demonstrate an approach to design topologically protected quantum circuits on the current generation of noisy quantum processors where spin-orbital coupling and related topological matter can be digitally simulated. In particular, a low-depth topological quantum circuit is performed on both IBM and Rigetti quantum processors. In the experiments, we not only observe but also distinguish the 0 and $\pi$ energy topological edge states by measuring qubit excitation distribution at the output of the circuits.

\end{abstract}

\maketitle

\emph{Introduction}.
Discoveries of topological insulators \cite{TI1,TI2} and the quantum Hall effect \cite{QH} have boosted the exploration of novel topological phases of matters, enabling fascinating physics research and exciting opportunities for new devices. Although topological quantum states were first discovered in solid-state materials, quantum simulators provide opportunities to go beyond what is possible in real materials, taking advantages of the high controllability and flexibility of these platforms \cite{CA1,CA2,Bloch,Blatt,Monroe,Koch,Martinis,Walther}. And such high tunability could greatly enhance the prospects of probing exotic topological phases \cite{TPCA1,TPCA2,TPCA3,Gadway2018,Wang2019,Yan2019,TPQW1,Xu2016,TPQW2,TPQW3,
TPQW4,TPQW5,MeiJin2019,Xu2018,TPQW6,TPQW7,TPQW8,MeiSun2019,Amin2018,Siddiqi2017,Roushan2017,Lehnert2014,Wang2016,
Roushan2014,Yin2018,ZhuYu2018,ZhuPan2018,WangYu2019}.

In stark contrast to analog quantum simulation that mimics the time evolution of one specific model Hamiltonian \cite{Lloyd,Nori}, digital quantum simulation (DQS) has the advantage of being universally applicable. This universality is a result of the fact that DQS encodes the state of the quantum systems onto qubits and emulates the time evolution through repeated cycles of qubit rotations (quantum gates) by means of quantum algorithms \cite{Lloyd,Nori,Blatt2011,DQS3,DQS4,DQS5}. Such a circuit-based simulator can, in principle, efficiently simulate any finite-dimensional local Hamiltonian. It now has been widely applied to study quantum chemistry~\cite{QuanChem1,QuanChem2}, many-body models~\cite{Zoller2010,Solano2012a,Solano2012b,Solano2014,
Martinis2015,Wallraff2015,Martinis2016,Knolle2019}, high-energy physics~\cite{HEP1,HEP2,HEP3,HEP4,HEP5,HEP6,HEP7} and even geometric phases~\cite{GP1,GP2}. In regard to topology, digitally simulating a topological Hamiltonian has been reported in single-qubit parameter space~\cite{Du2014}. However, how to perform digital simulation of topological matter with a programmable quantum processor remains unknown.

In this Letter, we develop and demonstrate a unique technique of designing topologically protected quantum circuits on the current generation of noisy quantum processors. Spin-orbital coupling (SOC) and related topological matter can be digitally simulated in the protocol by employing common single- and two-qubit quantum gates, thus providing the essential component to generate various topological phases of matter. We also show that quantum circuits could be used to measure the hallmarks of topological matter, including topological invariants and edge states. Furthermore, we experimentally perform a low-depth topological quantum circuit to simulate one-dimensional (1D) topological phases on the currently available IBM and Rigetti quantum processors. Strikingly, by measuring the qubit excitation distribution at the output of the circuit, we observe and distinguish the 0 and $\pi$ energy topological edge states, which have not been previously reported~\cite{Api}. Finally, we apply our method to design a quantum circuit for DQS of 2D SOCs and anomalous topological Chern insulator phases. All of our results are tenable with noisy quantum circuits, which manifests the power of topological protection.

\begin{figure}[h]
\includegraphics[width=8cm,height=3cm]{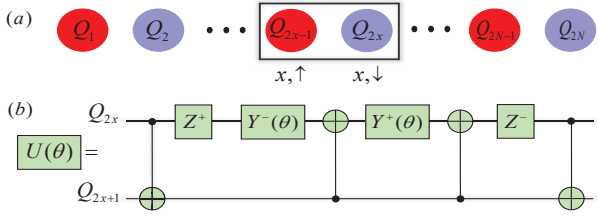}
\caption{(a) Mapping a 1D spinful lattice into qubits in a quantum circuit. A pair of qubits ($Q_{2x-1},Q_{2x}$) are employed to simulate a spin particle at a lattice site ($x_{\uparrow},x_{\downarrow}$). (b) Quantum circuit for implementing two-site SOC.}
\label{Fig1}
\end{figure}

\emph{Quantum circuits for SOC and the related topological matter}. In solid-state materials, SOC originates from the movement of electrons in a crystal's intrinsic electric field, and is one of the prominent mechanisms to induce the topological phases of matter~\cite{TI1,TI2,TPCA2,Galitski}. In our work, we shall present a unique approach to implement DQS of SOC using programmable quantum circuits. Before illustrating our procedure in detail, we summarize its underlying mechanism: a spinful lattice with $N$ sites is mapped into a qubit array with $2N$ qubits in a quantum circuit. At each site $x$, a spin particle is emulated by two qubits ($Q_{2x-1}$, $Q_{2x}$), representing spin up and spin down, respectively. By judiciously designing the gate sequences, we can digitally simulate the on-site spin rotation and two-site SOC, which are the basic buliding blocks of our approach. Finally, by means of Floquet sequences including these two circuit-based building blocks, we can simulate all forms of SOC \cite{Galitski}, which play essential roles in driving the phases to be topologically nontrivial.

As shown in Fig. \ref{Fig1}, we map a spinful lattice into a qubit array. The ground and excited states of each qubit are labelled as $|g\rangle$ and $|e\rangle$ respectively. When all qubits are prepared in the ground state, the state of the system is corresponding to the vacuum state of the simulated lattice $|\text{vac}\rangle=|gg\cdot\cdot\cdot gg\rangle$. Flipping one of the qubits  into the excited state $|e\rangle$ simulates the creation of a particle in the lattice. Specifically, exciting the odd (even) qubit $Q_{2x-1}$ ($Q_{2x}$) creates a spin up (down) particle at the lattice site $x$. The associated raising operators are mapped into hard-core bosonic creation operators through
$a^{\dag}_{x,\uparrow}=|e\rangle_{Q_{2x-1}}\langle g|$ and $a^{\dag}_{x,\downarrow}=|e\rangle_{Q_{2x}}\langle g|$.
Therefore, the coupling between the qubits $Q_{2x-1}$ and $Q_{2x}$ simulates the on-site spin rotation
\begin{equation}
\tilde{H}_o(x)=J_oa^{\dag}_{x,\uparrow}a_{x,\downarrow}+H.c.,
\end{equation}
while the coupling between the qubits $Q_{2x}$ and $Q_{2x+1}$ simulates the two-site SOC
\begin{equation}
\tilde{H}_s(x)=J_sa^{\dag}_{x,\downarrow}a_{x+1,\uparrow}+H.c..
\end{equation}
Here we will show that the dynanmics of the SOC and on-site spin rotations can be exactly programmed on a quantum circuit consisting of sequences of universal quantum gates, including single-qubit rotation gates and two-qubit entangled gates. We now address our protocol to programme the time evolution of the two-site SOC $\tilde{H}_s$ with two- and single-qubit quantum gates acting on the qubits ($Q_{2x},Q_{2x+1}$). After an evolution time $t$, the evolution operator is expressed as $U(t)=e^{-i\tilde{H}_st}$. Note that the excitation number in the qubit array is conserved, i.e., the time evolution only works in the single excitation subspace if the initial input state of the circuit is a single excitation state. The single excitation subspace for the two qubits ($Q_{2x},Q_{2x+1}$) is $\{|ge\rangle,|eg\rangle\}$. In this subspace, the evolution operator can be rewritten as
\begin{align}
U(\theta)=\cos(\theta)(|ge\rangle_{Q_{2x},Q_{2x+1}}\langle ge|+|eg\rangle_{Q_{2x},Q_{2x+1}}\langle eg|) \nonumber\\
-i\sin(\theta)(|ge\rangle_{Q_{2x},Q_{2x+1}}\langle eg|+|eg\rangle_{Q_{2x},Q_{2x+1}}\langle ge|),\nonumber
\end{align}
where $\theta=J_st$. We find that such a time evolution operation can be precisely programmed through a sequence of four two-qubit controlled-NOT (CNOT) gates and four single-qubit rotation gates, i.e.,
\begin{align}
U(\theta)&=\text{CNOT}_{Q_{2x},Q_{2x+1}}\cdot Z^{+}_{Q_{2x}}\cdot Y^{-}_{Q_{2x}}(\theta)\cdot\text{CNOT}_{Q_{2x+1},Q_{2x}}
\nonumber \\
&\cdot Y^{+}_{Q_{2x}}(\theta)\cdot\text{CNOT}_{Q_{2x+1},Q_{2x}}\cdot Z^{-}_{Q_{2x}}\cdot\text{CNOT}_{Q_{2x},Q_{2x+1}},\nonumber
\end{align}
where $Y^{\pm}(\theta)=e^{\mp i\frac{\theta}{2}\sigma_y}$ and $Z^{\pm}=e^{\mp i\frac{3\pi}{4}\sigma_z}$. 
The corresponding quantum circuit is shown in Fig. \ref{Fig1}(b). One can easily demonstrate that this circuit acts only on the single excitation qubit states $\{|ge\rangle,|eg\rangle\}$, leaving the other two states $\{|gg\rangle,|ee\rangle\}$ unchanged. Similarly, the time evolution governed by the on-site Hamiltonian $\tilde{H}_o(x)$ can be programmed through the same form of composite quantum gate $U$ acting on two qubits ($Q_{2x-1},Q_{2x}$).

\begin{figure}[t]
\includegraphics[width=8cm,height=8cm]{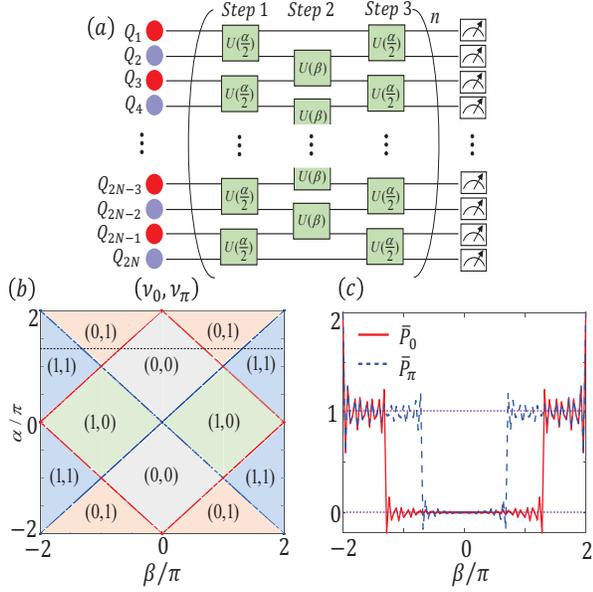}
\caption{(a) Quantum circuits for 1D topological insulator. The basic unit gates consist of three steps of composite two-qubit gates $U$, realizing DQS of an effective Hamiltonian $H_1$. $n$ is the cycle number of the unit gates. (b) The phase diagram of the quantum circuits varies with rotation angles ($\alpha$, $\beta$) in the single-qubit quantum gates. (c) By measuring the centers of the mean displacements ($\bar{P}_{0}$, $\bar{P}_{\pi}$), one can detect the winding numbers ($\nu_0$, $\nu_{\pi}$) as a function of $\beta$ for $\alpha=1.3\pi$ (the dashed line shown in (b)).}
\label{Fig2}
\end{figure}

To digitally simulate the topological quantum matter described by an effective Hamiltonian $H_{eff}$ consisting of $\tilde{H}_o(x)$ and $\tilde{H}_s(x)$, one can construct a quantum circuit based on the aforementioned programmable pairwise SOC and on-site spin rotations to mimick  the evolution operation
\begin{equation}
U_T=e^{-iH_{eff} T},
\label{FlOp}
\end{equation}
where $T$ is a time period. We will illustrate this approach with typical examples in 1D and 2D topological matter. The generalization of the method to higher dimensions is straightforward.

\emph{Quantum circuits for 1D topological matter}. We illustrate the approach by simulating a 1D topological Floquet phase described by the time evolution operator $U_1\equiv e^{-iH_1 T}$ with the effective Hamiltonian $H_1$ given by Eq. (S6) in Supplemental Material (SM) \cite{SM}. We find that the $U_1$ can be decomposed as
 \begin{equation}
U_{1}\equiv e^{-iH_1 T}=e^{-iH_oT/3}e^{-iH_sT/3}e^{-iH_oT/3},
\label{DQSU1}
\end{equation}
where $H_o=\sum_x\tilde{H}_o(x)$ is the on-site spin rotation and $H_s=\sum_x\tilde{H}_s(x)$ is the 1D SOC. Based on Eq.(\ref{DQSU1}), we can construct a quantum circuit with three steps of parallel two-qubit composite quantum gates $U$ to simulate the evolution $U_1$, i.e,
\begin{align}
U_1&=\bigotimes^N_{x=1}U(\frac{\alpha}{2})_{Q_{2x-1},Q_{2x}}
\cdot\bigotimes^{N-1}_{x=1}U(\beta)_{Q_{2x},Q_{2x+1}}\nonumber \\
&\cdot\bigotimes^N_{x=1}U(\frac{\alpha}{2})_{Q_{2x-1},Q_{2x}},
\label{U1gate}
\end{align}
where $U(\eta)_{Q_{m},Q_{n}}$ denotes the quantum gate $U(\eta)$ acting on the qubits $Q_m$ and $Q_n$ with the parameters $J_o=3\alpha/4T$ and $J_s=3\beta/2T$, as illustrated in Fig. \ref{Fig2}(a). Here the first (third) and second steps simulate the time evolution of pairwise on-site spin rotations and pairwise two-site SOCs, respectively.

The topology of the $U_1$-based quantum circuit is characterized by the topological winding number~\cite{Rudner,Michel,Delplace}
\begin{equation}
\nu_{\epsilon}=\frac{i}{4\pi}\int^{\pi}_{-\pi}dk_x\text{tr}
\big(\tau_zU^{-1}_{1\epsilon}\partial_{k_x}U_{1\epsilon}\big),
\end{equation}
where $U_{1\epsilon}$ ($\epsilon=0,\pi$) is the periodized evolution operator defined by $U_1$ in the momentum space, see SM \cite{SM}. The values of $\nu_0$ and $\nu_{\pi}$ are numerically calculated and the results are plotted in Fig. \ref{Fig2}(b). By changing the rotation angles ($\alpha$, $\beta$), we can tune the quantum circuit into various topological phases characterized by $\nu_{0}=0,1$ and $\nu_{\pi}=0,1$. Interestingly, by measuring mean displacements~\cite{MD,TPQW5}, we find that the winding numbers ($\nu_0, \nu_{\pi}$) can be directly detected via two quantum circuits $(U_1)^n$ and $(U^{\prime}_1)^n$~\cite{SM}, where the starting point in $U^{\prime}_1$ is the composite gate $U(\beta)$ instead of $U(\frac{\alpha}{2})$. Suppose the input state for both circuits is $|\psi_{in}\rangle=|gg...e...gg\rangle$. We measure the mean displacement $\bar{P}^{\prime}=\langle\psi^{\prime}_{out}|\sum_{x=1}^{N}x|e\rangle _{Q_{2x}}\langle e|)|\psi^{\prime}_{out}\rangle$ at the output of the circuit, where $|\psi^{\prime}_{out}\rangle=(U^{\prime}_1)^n|\psi_{in}\rangle$. As shown in Fig. \ref{Fig2}(c), the values of the winding numbers ($\nu_0, \nu_{\pi}$) are equal to the oscillation center of the mean displacements ($\bar{P}_{0}=-\bar{P}-\bar{P}^{\prime}$, $\bar{P}_{\pi}=\bar{P}-\bar{P}^{\prime}$), which can be derived from the even qubit excitation distribution at the output of the circuit~\cite{SM}.

\emph{Experimental implementation on quantum processors}. Free access IBM~\cite{IBM} and Rigetti~\cite{Rigetti} quantum processors have recently been successfully applied to study many-body quantum states~\cite{Chiesa2019,Motta2019,Choo2018,Smith2019}. Here we perform a topological quantum circuit on both IBM and Rigetti quantum processors. Specifically, an eight-qubit quantum circuit $(U_1)^3$ is programmed on \emph{ibm}$\_$16$\_$\emph{melbourne}~\cite{SM}. In our quantum circuits, topological edge states, the hallmark of topological phases, appear at both $E=0$ and $\pi$, with the number of modes determined by $\nu_0$ and $\nu_{\pi}$ respectively~\cite{SM}.

In SM ~\cite{SM}, we  analytically derive the wave functions of the edge states associated with the Floquet topological phases. In particular, the wave functions of the topological edge states localized at the left boundary with energy $E=0$ or $\pi$ are derived as
\begin{align}
|\psi^0_L\rangle&=\sum^{N}_{x=1}{\lambda_1}^{x-1}|e\rangle_{Q_{2x-1}},\,\,
|\psi^{\pi}_L\rangle&=\sum^{N}_{x=1}{\tilde{\lambda}_2}^{x-1}|e\rangle_{Q_{2x}},\nonumber
\end{align}
where $\lambda_1=-\tan(\frac{\alpha}{4})\cot(\frac{\beta}{4})$ and $\tilde{\lambda}_2=\cot(\frac{\alpha}{4})\cot(\frac{\beta}{4})$.
Interestingly, the probability of qubit excitations in $|\psi^{0(\pi)}_L\rangle$ is maximized in the leftmost odd (even) qubit $Q_1$ ($Q_2$). This feature allows us to experimentally observe and discriminate the $0$ and $\pi$ energy edge states.

\begin{figure}[h]
\includegraphics[width=9cm,height=6cm]{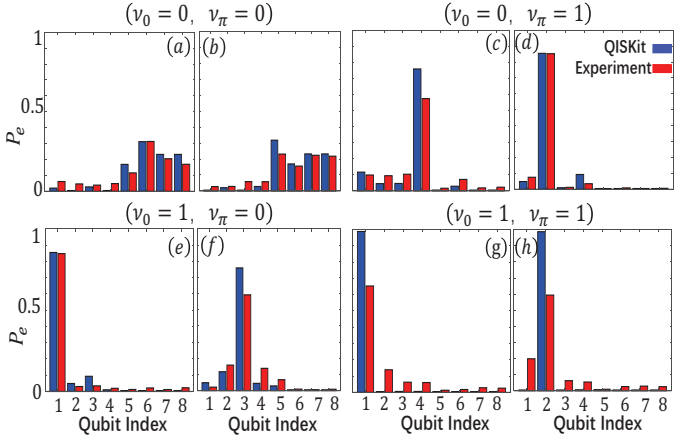}
\caption{Experimental observation of 0 and $\pi$ energy topological edge states on the IBM quantum processor. The qubit excitation probability distributions are measured at the output of an eight-qubit topological quantum circuit when the input qubit states are (a,c,e,g) $|\psi^{0}_{\text{in}}\rangle$ and (b,d,f,h) $|\psi^{\pi}_{\text{in}}\rangle$. The single-qubit rotation angles are chosen as (a,b) $\alpha=\pi$ and $\beta=0.9\pi$; (c,d) $\alpha=1.9\pi$ and $\beta=0.4\pi$; (e,f) $\alpha=0.1\pi$ and $\beta=0.7\pi$; (g,h) $\alpha=\pi$ and $\beta=1.9\pi$. }
\label{Fig3}
\end{figure}

To observe the edge states $|\psi^{0(\pi)}_L\rangle$, we prepare the initial input state of the circuit at $|\psi^{0(\pi)}_{\text{in}}\rangle=|eggggggg\rangle(|gegggggg\rangle)$, which yields the output state $|\psi^{0(\pi)}_{\text{out}}\rangle=(U_1)^3|\psi^{0(\pi)}_{\text{in}}\rangle$. The measured qubit excitation distributions at the output acquired from the IBM quantum processor are shown in Fig. \ref{Fig3}. When the quantum circuit is tuned into the topological phase with $\nu_{0}=1$ ($\nu_{\pi}=1$) and supports one 0 ($\pi$) energy edge state, as shown in Figs. \ref{Fig3}(e) and \ref{Fig3}(g) (Figs. \ref{Fig3}(d) and \ref{Fig3}(h)), the qubit excitation at the output maximally populates the leftmost odd (even) qubit $Q_1$ ($Q_2$). The reason for this is that the initial input state $|\psi^{0(\pi)}_{\text{in}}\rangle$ has a large overlap with $|\psi^{0(\pi)}_L\rangle$, while the qubit excitation evolves mainly in the circuit based on the $0$ ($\pi$) energy edge state wave packet and always maximally localizes in the leftmost odd (even) qubit $Q_1$ ($Q_2$). In contrast, as shown in Figs. \ref{Fig3}(a) and \ref{Fig3}(c) (Figs. \ref{Fig3}(b) and \ref{Fig3}(f)), when the quantum circuits are tuned into the trivial phases with $\nu_0=0$ ($\nu_{\pi}=0$) and there is no overlap with the edge states $|\psi^{0(\pi)}_L\rangle$, the qubit excitations are transferred into the bulk of the circuit and do not maximally populate the leftmost edge qubits. This is because in this case the initial state is a superposition of different Bloch bulk states and the qubit excitations transfer in the circuit via bulk state wave packets. Therefore, the 0 and $\pi$ energy edge states can be distinguished by observing whether the leftmost odd qubit $Q_1$ or the leftmost even qubit $Q_2$ is maximally excited. The experimental results acquired from the Rigetti quantum processor are shown in SM ~\cite{SM} and demonstrate similar results. In Fig. \ref{Fig3}, we show the simulation results derived from noisy quantum circuit QISKit. The mismatch between the experimental and the simulated results is due to the errors caused by the crosstalk in the quantum gates and the imperfections in state readouts~\cite{SM}. However, due to topological protection, edge state localization can still be unambiguously observed in noisy quantum circuits.

\begin{figure}[h]
\includegraphics[width=9cm,height=7cm]{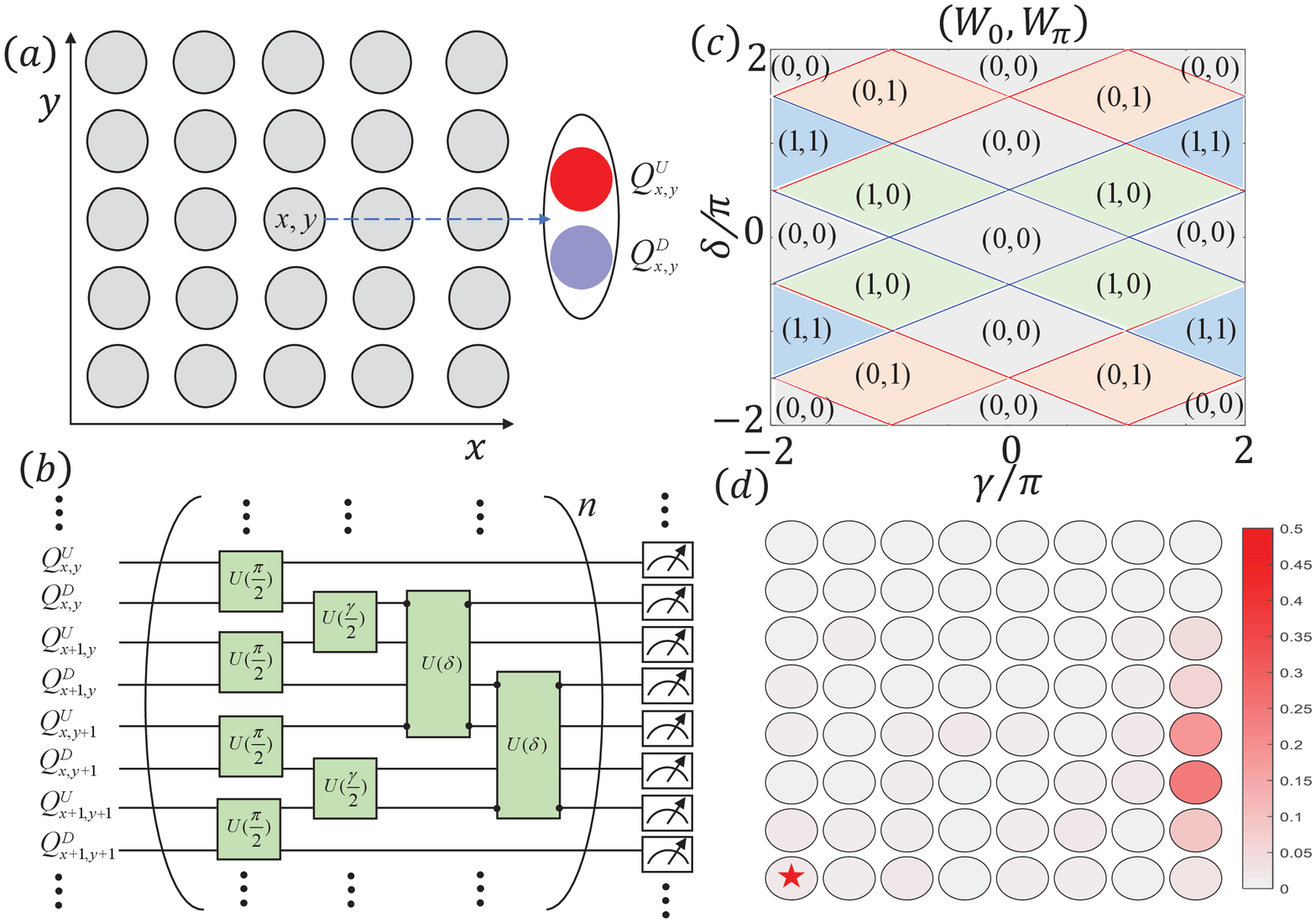}
\caption{(a) Simulating a 2D spinful lattice with  a quantum circuit. Each lattice site is encoded by two qubits $Q^U_{x,y}$ and $Q^D_{x,y}$. (b) A quantum circuit for 2D topological matter. (c) The  Floquet winding numbers $(W_0,W_{\pi})$ as a function of single-qubit rotation angles ($\gamma$, $\delta$). (d) Qubit excitation distribution at the output of a circuit with ($\gamma=1.9\pi,\delta=0.8\pi$) and $n=10$ after the qubit $Q^U_{x=1,y=1}$ (labelled by the star) is excited.}
\label{Fig4}
\end{figure}

\emph{Quantum circuits for 2D topological matter}. Our approach is universal and can be applied in high-dimensional systems. In the following example, we show how to design quantum circuits protected by 2D topology. As with the 1D case, we exploit two qubits $Q^U_{x,y}$ and $Q^D_{x,y}$ to simulate a site of a 2D spinful lattice (see Fig. \ref{Fig4}(a)), where the raising operators $|e\rangle_{Q^U_{x,y}}\langle g|$ and $|e\rangle_{Q^D_{x,y}}\langle g|$ are mapped to the creation operators $a^{\dag}_{x,y,\uparrow}$ and $a^{\dag}_{x,y,\downarrow}$ respectively.
 We aim to simulate a 2D topological Floquet phase described by the time evolution operator $U_2\equiv e^{-iH_2 T}$ with the effective Hamiltonian $H_2$ given by Eq. (S66) in SM \cite{SM}. We find that the $U_2$ can be decomposed as
 \begin{equation}
U_{2}=e^{-iH_{sy}T/3}e^{-iH_{sx}T/3}e^{-iH_{os}T/3},\label{DQSU2}
\end{equation}
where $H_{os}=\sum_{x,y}\tilde{H}_o(x,y)$ with $\tilde{H}_o(x,y)=J_{os}a^{\dag}_{x,y,\uparrow}a_{x,y,\downarrow}+H.c.$ and $J_{os}=3\pi/4T$, $H_{sx}=\sum_{x,y}\tilde{H}_{sx}(x,y)$ with $\tilde{H}_{sx}(x,y)=J_{sx}a^{\dag}_{x,y,\downarrow}a_{x+1,y,\uparrow}+H.c$  and  $J_{sx}=3\gamma/4T$, $H_{sy}=\sum_{x,y}\tilde{H}_{sy}(x,y)$ with $\tilde{H}_{sy}(x,y)=J_{sy} a^{\dag}_{x,y,\downarrow}a_{x,y+1,\uparrow}+H.c.$ and $J_{sy}=3\delta/2T$. As shown in Fig. \ref{Fig4}(b), we further find that such a circuit can be constructed as
\begin{align}
U_2&=\bigotimes^{N-1}_{x=1}U(\delta)_{Q^D_{x,y},Q^U_{x,y+1}}
\cdot\bigotimes^{N-1}_{x=1}U(\frac{\gamma}{2})_{Q^D_{x,y},Q^U_{x+1,y}}\nonumber \\
&\cdot\bigotimes^N_{x=1}U(\frac{\pi}{2})_{Q^U_{x,y},Q^D_{x,y}},
\label{U2gate}
\end{align}
where the quantum gate $U(\frac{\pi}{2})_{Q^D_{x,y},Q^U_{x,y}}$ in the first step programs the time evolution of on-site spin rotation, and the gates $U(\frac{\gamma}{2})_{Q^D_{x,y},Q^U_{x+1,y}}$ and $U(\delta)_{Q^D_{x,y},Q^U_{x,y+1}}$ in the second and third steps program the time evolution of x- and y-direction SOCs, respectively.


The topology of the $U_2$-based quantum circuit is characterized by the Floquet winding number $W_{\epsilon=0,\pi}$ ~\cite{Demler,Rudner,Mueller}. In Fig. \ref{Fig4}(c), we show the rich topological phase diagram versus  single-qubit rotation angles $(\gamma,\delta)$. There are four 2D Floquet topological phases featured by $W_0=0,1$ and $W_{\pi}=0,1$. Although the Chern number for the Floquet topological phase $(W_0=1,W_{\pi}=1)$ is zero, the system supports topological edge states~\cite{SM}, which can be used as topologically protected quantum channels to transfer quantum state~\cite{Lukin,Zoller,Mei2018,Gong2019}. We assume that  the qubit $Q^U_{x=1,y=1}$ is initially excited, as shown in Fig. \ref{Fig4}(b). We use noisy quantum circuit QISKit to calculate qubit excitation distribution at the output~\cite{SM} and the results are plotted in Fig. \ref{Fig4}(d). It turns out that the qubit excitation state can still be transferred along the edges via the edge state quantum channel, which reflects the power of quantum circuits with topological protection.

In summary, we have theoretically developed and experimentally demonstrated an approach to implement DQS of SOCs and the associated topological phases of matter on current generation of noisy intermediate-scale quantum processors. Our approach is generic and can be used to simulate a wide variety of complex SOCs, such as Rashba \cite{Rashba} and Dresselhaus SOC \cite{Dresselhau}, and related topological phases that are intractable in atomic, photonic, and solid-state systems. By applying the proposed method and utilizing current quantum processors, we can, in principle, build and explore the topological matters we are interested in, such as high-dimensional and high-order topological matters that are currently highly pursued \cite{TI1,TI2,HOTI}. When multiple qubit excitations are put into topological quantum circuits, interacting topological states \cite{STP} also can be investigated on noisy intermediate-scale quantum processors, which are beyond the computation capability of current classical computers and expected to show quantum advantages. In addition, it would be interesting to investigate topological quantum information processing tasks via topological quantum circuits~\cite{TPQIP,TPQC1,TPQC2}.

\begin{acknowledgments}
\emph{Acknowledgements}. F. M. thanks Heng Shen for invaluable help in improving the presentation of our manuscript and thanks Ying Hu and Dong-Ling Deng for helpful discussions. This work was supported by the National Key Research and Development Program of China
(2017YFA0304203), National Natural Science Foundation of China (NSFC)
(11604392), Changjiang Scholars and Innovative Research Team in University of Ministry of Education of China (PCSIRT)(IRT\_17R70), Fund for Shanxi 1331 Project Key Subjects Construction, and 111 Project (D18001). Y.F. Yu was supported by National Natural Science Foundation of China (Nos.61775062). S. L. Z. was supported by the Key-Area Research and Development Program of GuangDong Province (Grant No. 2019B030330001), the National Key Research and Development Program of China (Grant No.
2016YFA0301800), the National Natural Science Foundation of China (Grant No. 91636218), and the Key Project of Science and Technology of Guangzhou (Grant No. 201804020055).
\end{acknowledgments}

\end{document}